\documentclass[12pt]{article}
\usepackage{amssymb, amsmath, amsfonts,amstext, xspace, lscape,  latexsym, 
diagrams}
\begin{document}

\newtheorem{Th}{Theorem}[section]
\newtheorem{Def}{Definition}[section]
\newtheorem{Lemm}{Lemma}[section]
\newtheorem{Cor}{Corollary}[section]
\newtheorem{Rem}{Remark}[section]
\newtheorem{Prop}{Proposition}[section]
\newtheorem{Reminder}{Reminder}[section]
\newtheorem{Conj}{Conjecture}

\title{\bf Exact Diagonalisation of The XY-Hamiltonian of Open Linear 
Chains with Periodic Coupling Constants and Its Application to Dynamics of 
One-Dimensional Spin Systems}
\author{K.~E.~Feldman\footnote{E-mail: k.feldman@dpmms.cam.ac.uk}}
\date{}
\markboth{}{Exact Solution of Open Linear Chains with Periodic Coupling 
Constants}
\maketitle
\begin{center}
\footnotesize{{\it DPMMS, CMS,  University of Cambridge, Wilberforce Road, 
Cambridge, CB3 0WB, UK}}
\end{center}
\begin{abstract}
A new method of diagonalisation of the XY-Hamiltonian of inhomogeneous open 
linear chains with periodic (in space) changing Larmor frequencies and 
coupling constants is developed. As an example of application, multiple 
quantum dynamics of an inhomogeneous chain consisting of 1001 spins is 
investigated. Intensities of multiple quantum coherences are calculated for 
arbitrary inhomogeneous chains in the approximation of the next nearest 
interactions.


\end{abstract}

\section{Introduction}

One-dimensional exactly solvable models (spin chains, rings)~\cite{Gaudin} 
have been actively employed for studying various problems of spin 
dynamics~\cite{FeldLac} and quantum information theory~\cite{NielsenChuang}. 
Substantial progress in our understanding of spin dynamics has been 
achieved on the basis of a homogeneous  XY model for spin chains ($s=1/2$) in 
transverse magnetic field~\cite{Lieb}. Recently, Hamiltonians of the simplest 
inhomogeneous systems (alternating systems) have been diagonalised for the 
ring~\cite{Ye} and the linear spin chain~\cite{FeldmanRud}. Development of 
methods for exact solution of inhomogeneous spin problems has become 
especially urgent in conjunction with recent progress in quantum computation 
and quantum information theories~\cite{NielsenChuang}. 
In particular, these methods can be used for studying the quantum state 
transfer from one end of the chain to another one~\cite{CDEL,ACDE}
The qubit addressing problem can be attacked by variation of the Larmor 
frequencies of different spins~\cite{FeldDor} in classical one-dimensional 
models. This immediately brings inhomogeneity into the XY model as diagonal 
elements of the corresponding Hamiltonian are not equal. Consequently, we 
arrive at one-dimensional spin models with Hamiltonians described by 
three-diagonal matrices which elements on the diagonal are not-equal, and  
those under and over the main diagonal are not identically equal as well. 
Low sensitivity of nuclear magnetic resonance (NMR), which is widely applied 
in experimental implementation of quantum computations~\cite{NielsenChuang}, 
leads to a further complication of the model described above. Consideration of 
$k$-qubit systems brings us naturally to the study of $kn+r$ spin chains 
($r<k$, $n$ is arbitrary, $k,n,r$ are positive integers),
with Larmor frequencies and constants of spin-spin interactions repeating 
periodically with period $k$.

The paper suggests a new method of diagonalisation of a Hamiltonian of 
a linear $k$-periodic spin chain of length $kn+r$ for various values $r<k$. 
Special attention is given to the case $k=3$. For a spin chain of $3n+2$ sites 
with periodic parameters (period 3) a multiple quantum 
dynamics~\cite{Baum,FeldLac} is analysed and intensities of all MQ coherences 
are calculated (see Figure 2). This analysis is based on the explicit 
diagonalisation of the
Hamiltonian of a linear 3-periodic spin chain with $3n+2$ sites 
(Theorem~\ref{Th3n+2}) and the exact formulae, obtained in the paper 
(Theorem~\ref{Trace}), for the intensities of all MQ coherences developed in 
any nuclear spin system coupled by the nearest neighbour dipolar interactions. 
We conclude the paper 
with a discussion of the properties of eigenvalues and eigenvectors of the  
general $k$-periodic systems with $kn-1$ sites (Theorem~\ref{MAINkn-1}).


\section{Model}

The Hamiltonian of a spin-1/2 open chain with only nearest neighbour (NN)
couplings has the following general form
\begin{equation}
\label{TheorHam}
H=\sum^N_{n=1}\omega_n I_{nz}+\sum^{N-1}_{n=1} D_{n,n+1}
\left(I_{n,x} I_{n+1,x}+I_{n,y} I_{n+1,y}\right),
\end{equation}
where $\omega_n$, $n=1,...,N$, are the Larmor frequencies,  and 
$D_{n,n+1}$, $n=1,...,N-1$, are the NN coupling constants. The nuclear spins 
are specified by the spin-1/2 operators $I_{n\alpha}$ at the sites 
$n=1,...,N$ with the projections $\alpha=x,y,z$. The Jordan--Wigner 
transformation~\cite{Lieb}
\begin{equation}
\label{JW}
\begin{array}{l}
I_{n,-}=I_{n,x}-iI_{n,y}=(-2)^{n-1}\left(\prod^{l=n-1}_{l=1} 
I_{l,z}\right)c_n,\\
I_{n,+}=I_{n,x}+iI_{n,y}=(-2)^{n-1}\left(\prod^{l=n-1}_{l=1} 
I_{l,z}\right)c^+_n,\\
I_{n,z}=c^+_nc_n-1/2\\
\end{array}
\end{equation}
from the spin-1/2 operators $I_{n\alpha}$ to the creation (annihilation) 
operators $c^+_n$ $(c_n)$ of the spineless fermions takes the 
Hamiltonian~(\ref{TheorHam}) into the Hamiltonian
\begin{equation}
\label{JWN}
H=\sum^N_{n=1}\omega_n\left(c^+_nc_n-1/2\right)+
\frac{1}{2}
\sum^{N-1}_{n=1} D_{n,n+1}\left(c^+_nc_{n+1}+c^+_{n+1}c_n\right),
\end{equation} 
or in the matrix notations into
\begin{equation}
\label{JWM}
H=\frac{1}{2}{\bf c}^{+}\left(D+2\Omega\right){\bf c}-
\frac{1}{2}\sum^N_{n=1}\omega_n.
\end{equation}
In Eq.~(\ref{JWM}) we denote the row vector $\left(c^+_1,...,c^+_N\right)$ by
${\bf c}^+$, the column vector $\left(c_1,...,c_N\right)^t$ by
${\bf c}$ (the superscript '$t$' represents the transpose), and specify the 
matrices $\Omega$ and $D$ as
\begin{equation}
\label{GenMat}
\Omega=
\left[\begin{array}{ccccc}
\omega_1& 0&\cdots & 0& 0\\
0& \omega_1&\cdots & 0& 0\\
\vdots&\vdots&\ddots&\vdots&\vdots\\
0& 0&\cdots & \omega_{N-1}& 0\\
0& 0&\cdots & 0& \omega_N\\
\end{array}
\right],\quad
D=
\left[\begin{array}{ccccc}
0& D_1&\cdots & 0& 0\\
D_1& 0&\cdots & 0& 0\\
\vdots&\vdots&\ddots&\vdots&\vdots\\
0& 0&\cdots & 0& D_{N-1}\\
0& 0&\cdots & D_{N-1}& 0\\
\end{array}
\right].
\end{equation}
To diagonalise the matrix $D+2\Omega$ means to construct such unitary matrix 
$U=\{u_{nk}\}^N_{n,k=1}$ and diagonal matrix 
$\Lambda=diag\{\lambda_1,...,\lambda_N\}$ that
\begin{equation}
D+2\Omega=U\Lambda U^+,
\end{equation}
where the superscript '$+$' represents the conjugate transpose, and columns 
$(u_{1l},...,u_{Nl})^t$ of $U$, $l=1,...,N$, form an orthonormal basis of 
eigenvectors of $H=D+2\Omega$. 
The new fermion operators $\gamma^+_k$ and $\gamma_k$ introduced by the 
relations
\begin{equation}
c^+_n=\sum^N_{k=1}u^*_{nk}\gamma^+_k,\quad c_n=\sum^N_{k=1}u_{nk}\gamma_k
\end{equation}
bring Hamiltonian~(\ref{JWM}) into the Hamiltonian
\begin{equation}
H=\frac{1}{2}\sum^N_{k=1}\lambda_k\gamma^+_k\gamma_k-
\frac{1}{2}\sum^N_{n=1}\omega_n
\end{equation}
with energies $\lambda_{\nu}/2$ of the free fermion waves.

\

The problem of diagonalisation of $D+2\Omega$  with arbitrary constants
$\omega_n$, $n=1,...,N$, and $D_k$, $k=1,...,N-1$, for large $N$ 
($N\sim 10^6$) and computation of various functions of $D+2\Omega$ is a time
consuming problem usually dealt with the help of super computers. On the 
other hand, there are two known exact solutions (suitable for studying systems 
with large N) for such a problem when the spin-system considered 
has periodic changing Larmor frequencies and coupling constants. Namely, the 
case of equal sites 
($\omega_1=...=\omega_n=a$, $D_1=...=D_{N-1}=c$) has been solved 
in~\cite{Lieb}, while the case of period 2 with odd $N$ 
($\omega_1=\omega_3=...=\omega_N$, $\omega_2=\omega_4=...=\omega_{N-1}$,
$D_1=D_3=...=D_{N-2}$, $D_2=D_4=...=D_{N-1}$) has 
been solved in~\cite{FeldmanRud}. The present paper concerns properties of
a general periodic chain. In the following sections we shall demonstrate 
how to reduce the diagonalisation problem of the Hamiltonian of a $k$-periodic
chain with $kn-1$ sites ($k>2,n>1$) to the problem of finding roots for  
explicitly given polynomials of degree less or equal $k$. Therefore, the 
methods developed are particularly useful when $n>>k$. 

In what follows we will utilise essentially the result on 
diagonalisation of the homogeneous (in other words, 1-periodic) 
chain~\cite{Lieb} which we state now.
  
\begin{Lemm}
\label{JLAMBDA}
Let $c\neq 0$, then the matrix  
\begin{equation}
J_n(a,c)=\left[\begin{array}{ccccccc}
a& c& 0& \cdots& 0& 0& 0\\
c& a& c& \cdots& 0& 0& 0\\
0& c& a& \cdots& 0& 0& 0\\
\vdots& \vdots& \vdots&\ddots& \vdots& \vdots& \vdots\\
0& 0& 0&\cdots& a& c& 0\\
0& 0& 0&\cdots& c& a& c\\
0& 0& 0&\cdots& 0& c& a\\
\end{array}
\right]
\end{equation}
has $n$-distinct eigenvalues
\begin{equation}
\lambda_j=a+2c\cdot cos\left(\frac{\pi j}{n+1}\right),\quad j=1,...,n,
\end{equation}
and the corresponding eigenvectors are of the form 
\begin{equation}
\vec x_j=\left(sin\left(\frac{\pi j}{n+1}\right),sin\left(\frac{2\pi j}{n+1}\right),...,
sin\left(\frac{n\pi j}{n+1}\right)\right).
\end{equation}
\end{Lemm}
{\bf Proof.} It is a straightforward verification.


\section{Reduction over the Period}

Let us consider an open periodic chain of $kn+d$ sites 
($k>d\ge 0$, $k>1$) with $k$-periodic non--zero NN coupling constants: 
$$
D_1=D_{k+1}=...=D_{kn+1},\quad\cdots\quad D_d=D_{k+d}=...=D_{kn+d}, 
$$
\begin{equation}
D_{d+1}=D_{k+d+1}=...=D_{k(n-1)+d+1},\quad\cdots\quad
D_k=D_{2k}=...=D_{kn},
\end{equation}
and $k$-periodic Larmor frequencies 
$$
\omega_1=\omega_{k+1}=...=\omega_{kn+1},\quad\cdots\quad 
\omega_d=\omega_{k+d}=...=\omega_{kn+d},
$$ 
\begin{equation}
\omega_{d+1}=\omega_{k+d+1}=...=\omega_{k(n-1)+d+1},\quad\cdots\quad 
\omega_k=\omega_{2k}=...=\omega_{kn},
\end{equation}
($D_i,\omega_j\in \mathbb R$, $D_i\neq 0$, $i,j=1,...,k$). We 
shall demonstrate how to reduce the diagonalisation problem for the 
Hamiltonian of such a system to the problem of diagonalising of a certain 
$k\times k$ block matrix which entries are matrices of dimension $m\times m$, 
$(m+1)\times (m+1)$, $m\times (m+1)$ and $(m+1)\times m$ ($m=n$ in Section 5 
and $m=n-1$ in Section 7). A particular case of 
this reduction will be used in the consecutive sections for studying 
$k$-periodic chains of length $kn-1$.
 
Before we proceed we make an agreement regarding our notations. If a Gothic 
letter is  used in the description of any matrix as its element, then
this means that the corresponding place in the matrix is a matrix.  The size 
and the structure of a matrix denoted by a Gothic letter should be clear or 
given in the context. Therefore, the original matrix is a block matrix. To 
underline that the matrix consists of blocks we shall often denote
such matrices in bold. We shall also denote by $I_m$
the $m\times m$ identity matrix.

The Hamiltonian of a $k$-periodic system with $kn+d$ sites has the following 
form:
\begin{equation}
H=D+2\Omega,
\end{equation}
where 
\begin{equation}
\Omega=
\left[\begin{array}{cccccccccccc}
\omega_1     &   0   &\cdots &0      &0      &0      &\cdots &0 &  0    &\cdots &0      &  0    \\
0       &  \omega_2  &\cdots &0      &0      &0      &\cdots &0 &  0    &\cdots &0      &  0    \\
\vdots  &\vdots &\ddots &\vdots &\vdots &\vdots &\ddots &\vdots    &\vdots &\ddots &\vdots      &\vdots \\
0       &   0   &\cdots &\omega_{k-1}&0      & 0     &\cdots &0 &  0   &\cdots &0      &  0    \\
0       &   0   &\cdots &0      &\omega_k    & 0     &\cdots &0 &  0    &\cdots &0      &  0    \\
0       &   0   &\cdots &0      &0      & \omega_1   &\cdots &0 &  0    &\cdots &0      &  0    \\
 \vdots &\vdots &\ddots &\vdots &\vdots &\vdots &\ddots &\vdots   &\vdots &\ddots &\vdots &\vdots \\
0       &   0   &\cdots &0      &0      &0      &\cdots &\omega_k &  0    &\cdots &0      &  0    \\  
0       &   0   &\cdots &0      &0      &0      &\cdots &0   & \omega_1   &\cdots &0      &  0    \\
\vdots  &\vdots &\ddots &\vdots &\vdots &\vdots &\ddots &\vdots & \vdots &\ddots &\vdots &\vdots \\
0       &   0   &\cdots &0      &0      &0      &\cdots &0   & 0     &\cdots &\omega_{d-1}&  0  \\ 
0       &   0   &\cdots &0      &0      &0      &\cdots &0   & 0     &\cdots &0      &  \omega_d  \\
\end{array}\right]
\end{equation}
and
\begin{equation}
D=
\left[\begin{array}{cccccccccccc}
0       &   D_1 &0      &\cdots &0      &0      &0      &\cdots &  0    &\cdots &0      &  0    \\
D_1     &   0   &D_2    &\cdots &0      &0      &0      &\cdots &  0    &\cdots &0      &  0    \\
0       &   D_2 &0      &\cdots &0      &0      &0      &\cdots &  0    &\cdots &0      &  0    \\
\vdots  &\vdots &\vdots &\ddots &\vdots &\vdots &\vdots &\ddots &\vdots &\ddots &\vdots      &\vdots \\
0       &   0   &0      &\cdots &0      &D_{k-1}& 0     &\cdots &  0    &\cdots &0      &  0    \\
0       &   0   &0      &\cdots &D_{k-1}&0      &D_k    &\cdots &  0    &\cdots &0      &  0    \\
0       &   0   &0      &\cdots &0      &D_k    &0      &\cdots &  0    &\cdots &0      &  0    \\
 \vdots &\vdots &\vdots &\ddots &\vdots &\vdots &\vdots &\ddots &\vdots &\ddots &\vdots &\vdots \\  
0       &   0   &0      &\cdots &0      &0      &0      &\cdots &  0    &\cdots &0      &  0    \\
\vdots  &\vdots &\vdots &\ddots &\vdots &\vdots &\vdots &\ddots &\vdots &\ddots &\vdots &\vdots \\
0       &   0   &0      &\cdots &0      &0      &0      &\cdots & 0     &\cdots &0      & D_{d-1}  \\ 
0       &   0   &0      &\cdots &0      &0      &0      &\cdots & 0     &\cdots &D_{d-1}& 0  \\
\end{array}\right].
\end{equation}
To diagonalise matrix $H$ one finds (real) eigenvectors $u_{\nu}$ and 
(real) eigenvalues 
$\lambda_{\nu}$, $\nu=1$, ... $kn+d$, 
which satisfy the following equation
\begin{equation}
\label{EQ}
(D+2\Omega)u_{\nu}=\lambda_{\nu} u_{\nu}.
\end{equation}
To resolve this equation we associate to each vector $u\in {\mathbb R}^{kn+d}$ 
vectors $u_{(j)}$, $j=1,...,k$, formed by those
coordinates of $u$ whose numbers have residue $j$ modulo $k$. Observe that 
among vectors $u_{(1)},...,u_{(k)}$ there are
$d$-vectors of dimension $(n+1)$ and $(k-d)$ vectors of dimension $n$.  
Now equation~(\ref{EQ}) can be rewritten as a system
of $n$ linear equations in $u_{(j)}$:
\begin{equation}
\label{EQRED}
{\bf{\mathrm {\bf H}}}\left[\begin{array}{c}
u_{(1)}\\
\vdots\\
u_{(k)}\\
\end{array}\right]=\lambda_{\nu}\left[\begin{array}{c}
u_{(1)}\\
\vdots\\
u_{(k)}\\
\end{array}\right]
\end{equation}
with matrix 
${\bf{\mathrm {\bf H}}}=2{\bf{\mathrm{\bf \Omega}}}+{\bf {\mathrm  {\bf D}}}$.
Here
\begin{equation}
{\bf{\mathrm {\bf \Omega}}}=
\left[\begin{array}{cccccc}
{\cal W}_1&0&0&\cdots&0&0\\
0&{\cal W}_2&0&\cdots&0&0\\
0&0&{\cal W}_3&\cdots&0&0\\
\vdots&\vdots&\vdots&\ddots&\vdots&\vdots\\
0&0&0&\cdots&{\cal W}_{k-1}&0\\
0&0&0&\cdots&0&{\cal W}_k\\
\end{array}\right],
\end{equation}
${\cal W}_j=\omega_jI_{n+1}$ for $j=1,...,d$, 
${\cal W}_j=\omega_j I_n$ for $j=d+1,...,k$, 
and with an exception of two degenerate cases 
\begin{equation}
{\bf{\mathrm {\bf D}}}=\left[\begin{array}{ccccccccccc}
0      & {\cal D}_1   &0      &\cdots     &0              &0            &0      &0      &\cdots&0      &{\cal D}^t_k\\
{\cal D}_1    & 0     &{\cal D}_2    &\cdots     &0              &0            &0      &0      &\cdots&0      &0\\
0      & {\cal D}_2   &0      &\cdots     &0              &0            &0      &0      &\cdots&0      &0\\
\vdots &\vdots &\ddots &\ddots     &\vdots         &\vdots       &\vdots &\vdots &\ddots&\vdots &\vdots\\
0      &0      &\cdots &{\cal D}_{d-1}    &0              &{\cal D}^t_d &0      &0      &\cdots&0      &0\\
0      &0      &\cdots &0          &{\cal D}_d     &0            &{\cal D}_{d+1}&0      &\cdots&0      &0\\
0      &0      &0      &\cdots     &0              &{\cal D}_{d+1}      &0      &{\cal D}_{d+2}&\cdots&0      &0\\
\vdots &\vdots &\ddots &\ddots     &\vdots         &\vdots       &\ddots &\ddots &\ddots&\ddots &\vdots\\
0      &0      &0      &\cdots     &0              &0            &0      &\ddots      &\ddots&{\cal D}_{k-2}   &0\\
0      &0      &0      &\cdots     &0              &0            &0      &0      &\ddots&0      &{\cal D}_{k-1}\\
{\cal D}_k      &0      &0      &\cdots     &0              &0            &0      &0      &\cdots&{\cal D}_{k-1}&0\\
\end{array}\right],
\end{equation}
with ${\cal D}_j=D_jI_{n+1}$ for $j=1,2,...,d-1$, 
${\cal D}_j=D_jI_n$ for $j=d+1,...,k-1$, and 
${\cal D}_d,{\cal D}_k:{\mathbb R}^{n+1}\to {\mathbb R}^n$ given by
\begin{equation}
\label{DdDk}
{\cal D}_d=\left[\begin{array}{ccccccc}
D_d    &0      &0 &\cdots& 0&0&0\\
0      &D_d    &0 &\cdots& 0&0&0\\
0      &0      &D_d &\ddots& 0&0&0\\
\vdots &\vdots &\vdots&\ddots&\ddots&\vdots&\vdots\\
0&0&0&\cdots&D_d&0 &0\\
0&0&0&\cdots&0 &D_d&0\\
\end{array}\right],\qquad
{\cal D}_k=\left[\begin{array}{ccccccc}
0     &D_k  &0 &\cdots& 0&0&0\\
0     &0    &D_k &\cdots& 0&0&0\\
0     &0    &0 &\ddots& 0&0&0\\
\vdots&\vdots &\vdots&\ddots&\ddots&\vdots&\vdots\\
0&0&0&\cdots&0 &D_k &0\\
0&0&0&\cdots&0 &0   &D_k\\
\end{array}\right].
\end{equation}
It is easy to see that 
\begin{equation}
\label{relations}
{\cal D}_d{\cal D}^t_d=D^2_d I_n,\qquad {\cal D}_k{\cal D}^t_k=D^2_k I_n,\qquad
{\cal D}_d{\cal D}^t_k=D_dD_k J^t_n,\qquad {\cal D}_k{\cal D}^t_d=D_kD_d J_n,
\end{equation}
where $J_n$ is the Jordan $n\times n$ cell:
\begin{equation}
J_n=
\left[\begin{array}{cccccc}
0      & 1& 0   & \cdots &0      &0\\
0      & 0& 1   & \cdots &0      &0\\
\vdots & \ddots & \ddots &\ddots & \vdots &\vdots\\
0      & 0& 0   & \ddots &1      & 0\\
0      & 0& 0   & \cdots &0      & 1\\
0      & 0& 0   & \cdots &0      &0\\
\end{array}\right]. 
\end{equation}

There are two degenerate cases. In the case $k=2$ matrix ${\mathrm {\bf D}}$ 
decomposes as
\begin{equation}
{\mathrm {\bf D}}=
\left[\begin{array}{cc}
0& {\cal L}^t\\
{\cal L}& 0\\
\end{array}\right]
\end{equation}
where ${\cal L}: {\mathbb R}^{n+d}\to {\mathbb R}^n$:
\begin{equation}
{\cal L}=
\left[\begin{array}{ccccccc}
D_1    &D_2    &0   &\cdots& 0&0&0\\
0      &D_1    &D_2 &\cdots& 0&0&0\\
0      &0      &D_1 &\ddots& 0&0&0\\
\vdots &\vdots &\ddots&\ddots&\ddots&\vdots&\vdots\\
0      &0      &0&\ddots&\ddots&\ddots&\vdots\\
0      &0      &0&\cdots&0 &\ddots&\ddots\\
\end{array}\right],
\end{equation}
while in the case $k>2$ and $d=0$ we do not get matrix ${\cal D}_d$ and 
matrix ${\cal D}_k:\mathbb R^n\to\mathbb R^n$ is
of the form
\begin{equation}
{\cal D}_k=\left[\begin{array}{cccccc}
0     &D_k  &0 &\cdots& 0&0\\
0     &0    &D_k &\cdots& 0&0\\
0     &0    &0 &\ddots& 0&0\\
\vdots&\vdots &\vdots&\ddots&\ddots&\vdots\\
0&0&0&\cdots&0 &D_k\\
0&0&0&\cdots&0 &0  \\
\end{array}\right].
\end{equation}
It is easy to see that this matrix satisfies ${\cal D}_k{\cal D}^t_k=D^2_kI_{n-1,1}$ for a diagonal matrix $I_{n-1,1}$ with the first
$n-1$ diagonal elements equal to 1 and the last one equals to 0.

\

\begin{Rem}
From the reduction obtained above we divide consideration of the 
diagonalisation 
problem into three different cases: period $k=2,3$ and $k\ge 4$. In each case 
there are further reductions depending on the value of $d$. In the next 
sections we shall work out the case $k\ge 3$ and $d=k-1$.
\end{Rem}


\section{Some Auxiliary Results}

The diagonalisation process of Sections 5 and 7 will relay on some elementary 
facts about matrices of the form 
$H_{i,j}-\lambda I_{j-i+1}$ where for $i<j$:
\begin{equation}
H_{i,j}=
\left[\begin{array}{cccccc}
2\omega_i   & D_i & 0   &\cdots&0&0\\
D_i & 2\omega_{i+1}   & D_{i+1} &\cdots&0&0\\
0   & D_{i+1} & 2\omega_{i+2}   &\cdots&0&0\\
\vdots&\vdots&\vdots&\ddots&\vdots&\vdots\\
0&0&0&\cdots&2\omega_{j-1} &D_{j-1}\\
0&0&0&\cdots&D_{j-1}&2\omega_j\\
\end{array}\right].
\end{equation}
These facts are mostly know and we give their proofs only for the reader's 
convenience.

\begin{Lemm}
\label{distinct}
If $D_s\neq 0$ for all $s=i,...,j-1$, then all the eigenvalues of the matrix 
$H_{i,j}$ are distinct. 
\end{Lemm}
{\bf Proof.} Under the assumption that all $D_s$, $s=i,...,j$, are non--zero, 
one can reconstruct every eigenvector $u_{\nu}$ of $H_{i,j}$ by its eigenvalue 
$\lambda_{\nu}$ and the first coordinate $u_1$. Moreover, the expressions for 
all other coordinates $u_s$, $s=2,...,j-i+1$, of $u_{\nu}$ are linear in $u_1$.
This implies that every two eigenvectors $u_{\nu}$ and $u'_{\nu}$ with the 
same eigenvalue $\lambda_{\nu}$ are proportional with coefficient $u_1/u'_1$.

\begin{Lemm}
\label{adj}
Non-diagonal elements of the adjoint matrix of
$\left(H_{i,j}-\lambda I_{j-i+1}\right)$ are 
$$
\mathrm{adj}_{s,t}\left\{\left(H_{i,j}-\lambda I_{j-i+1}\right)\right\}=
\mathrm{adj}_{t,s}\left\{\left(H_{i,j}-\lambda I_{j-i+1}\right)\right\}=
$$
\begin{equation}
=(-1)^{s+t}det\left(H_{i,i+s-2}-\lambda I_{s-1}\right)
D_{i+s-1}\cdots D_{i+t-2}det\left(H_{i+t,j}-\lambda I_{j-i-t+1}\right)
\end{equation}
where the first index in $\mathrm{adj}_{s,t}\{\cdot\}$ denotes the row number, 
the second denotes the column number and $t>s$. The diagonal elements of the
adjoint matrix are
\begin{equation}
\mathrm{adj}_{s,s}\left\{\left(H_{i,j}-\lambda I_{j-i+1}\right)\right\}=
det\left(H_{i,i+s-1}-\lambda I_{s-1}\right)det\left(H_{i+s+1,j}-
\lambda I_{j-i-s}\right).
\end{equation}
\end{Lemm}
{\bf Proof.} The first part follows from an observation that element 
$D_{i+s-2}$ situated in row $(s-1)$ and column $s$ of matrix 
$\left(H_{i,j}-\lambda I_{j-i+1}\right)$ does not contribute to the 
cofactor $(s,t)$ of $\left(H_{i,j}-\lambda I_{j-i+1}\right)$, and similarly 
for element $D_{i+t-1}$ 
situated in row $t$ and column $(t+1)$. The second part is due to the 
splitting of the complement to the element $(s,s)$ of 
$\left(H_{i,j}-\lambda I_{j-i+1}\right)$ into 
$\left(H_{i,i+s-1}-\lambda I_{s-1}\right)$
and $\left(H_{i+s+1,j}-\lambda I_{j-i-s}\right)$.

\begin{Cor}
\label{INVERSE}
If we denote elements of $\left(H_{i,j}-\lambda I_{j-i+1}\right)^{-1}$ by $P_{s,t}$ ($s$ --- row, $t$ --- column), then
for $t>s$
\begin{equation}
P_{t,s}=P_{s,t}=(-1)^{s+t}\frac{det\left(H_{i,i+s-2}-\lambda I_{s-1}\right)D_{i+s-1}\cdots D_{i+t-2}det\left(H_{i+t,j}-\lambda I_{j-i-t+1}\right)}
{det\left(H_{i,j}-\lambda I_{j-i+1}\right)}.
\end{equation}
The diagonal terms $P_{t,t}$ are
\begin{equation}
P_{t,t}=\frac{det\left(H_{i,i+t-1}-\lambda I_{t-1}\right) 
det\left(H_{i+t+1,j}-\lambda I_{j-i-t}\right)}
{det\left(H_{i,j}-\lambda I_{j-i+1}\right)}.
\end{equation}
\end{Cor}

\begin{Lemm}
If $D_1\neq 0$, then
matrices $H_{1,k}$ and $H_{2,k}$ ($k\ge 2$) have no common eigenvalues.
\end{Lemm}
{\bf Proof.} First, observe that it is true for $k=2$:
\begin{equation}
det\left(H_{1,2}-\lambda I_2\right)=\left(2\omega_1-\lambda\right)\cdot
\left(2\omega_2-\lambda\right)-D^2_1,
\end{equation}
and if $\lambda$ is a root of $det\left(H_{2,2}-\lambda I_1\right)=0$,
then $\lambda=2\omega_2$, which is the root of
\begin{equation}
\left(2\omega_1-\lambda\right)\cdot
\left(2\omega_2-\lambda\right)-D^2_1=0
\end{equation}
only if $D_1=0$.

For general $k$, if $\lambda$ is a root of
\begin{equation}
det\left(H_{2,k}-\lambda I_{k-1}\right)=0,
\end{equation}
then for such  $\lambda$:
$$
det\left(H_{1,k}-\lambda I_k\right)=
(2\omega_1-\lambda)det\left(H_{2,k}-\lambda I_{k-1}\right)-
D^2_1det\left(H_{3,k}-\lambda I_{k-2}\right)=
$$
\begin{equation}
=-D^2_1det\left(H_{3,k}-\lambda I_{k-2}\right).
\end{equation}
For the last polynomial we can assume that it is non zero by inductive 
hypothesis and because $D_1\neq 0$. From here the statement follows.




\section{Exact Diagonalisation for a Spin Chain with $3n+2$ Sites}

To demonstrate how the reduction over the period works, first we study a 
partial case of the main result, namely a chain of period 3 with $3n+2$ sites. 
The reduction of Section 3 leads us to the following system of linear 
equations:
\begin{equation}
\label{3n+2}
\left[\begin{array}{ccc}
2{\cal W}_1-\lambda_{\nu}I_{n+1}& {\cal D}_1& {\cal D}^t_3\\
{\cal D}_1& 2{\cal W}_2-\lambda_{\nu}I_{n+1}& {\cal D}^t_2\\
{\cal D}_3& {\cal D}_2& 2{\cal W}_3-\lambda_{\nu}I_n\\
\end{array}\right]
\left[\begin{array}{c}
u_{(1)}\\
u_{(2)}\\
u_{(3)}\\
\end{array}\right]=0.
\end{equation}
Remind that ${\cal W}_j$, $j=1,2,3$, are just scalar matrices, as well as
the matrix ${\cal D}_1$, while  ${\cal D}_2,{\cal D}_3:
{\mathbb R}^{n+1}\to {\mathbb R}^n$ are given below:
\begin{equation}
\label{D2D3}
{\cal D}_2=\left[\begin{array}{ccccccc}
D_2    &0      &0 &\cdots& 0&0&0\\
0      &D_2    &0 &\cdots& 0&0&0\\
0      &0      &D_2 &\ddots& 0&0&0\\
\vdots &\vdots &\vdots&\ddots&\ddots&\vdots&\vdots\\
0&0&0&\cdots&D_2&0 &0\\
0&0&0&\cdots&0 &D_2&0\\
\end{array}\right],
\qquad
{\cal D}_3=\left[\begin{array}{ccccccc}
0     &D_3  &0 &\cdots& 0&0&0\\
0     &0    &D_3 &\cdots& 0&0&0\\
0     &0    &0 &\ddots& 0&0&0\\
\vdots&\vdots &\vdots&\ddots&\ddots&\vdots&\vdots\\
0&0&0&\cdots&0 &D_3 &0\\
0&0&0&\cdots&0 &0   &D_3\\
\end{array}\right].
\end{equation}
From the second equation of~(\ref{3n+2})
\begin{equation}
\left(\lambda_{\nu}-2\omega_2\right)\cdot u_{(2)}={\cal D}_1u_{(1)}+
{\cal D}^t_2u_{(3)}.
\end{equation}
Assume that an eigenvalue $\lambda_{\nu}$ of $H$ is not equal to $2w_2$,
then 
\begin{equation}
\label{u(2)}
u_{(2)}=\frac{D_1}{\lambda_{\nu}-2\omega_2}\cdot u_{(1)}+
\frac{1}{\lambda_{\nu}-2\omega_2}\cdot {\cal D}^t_2u_{(3)}.
\end{equation}
Substituting $u_{(2)}$ into the first equation and into the third equation 
of~(\ref{3n+2}), and using
that ${\cal D}_2{\cal D}^t_2=D^2_2I_n$ we obtain a system:
\begin{equation}
\label{Sys1}
\left\{
\begin{array}{l}
\left(D^2_1-(\lambda_{\nu}-2\omega_1)(\lambda_{\nu}-2\omega_2)\right)u_{(1)}
+\left(D_1{\cal D}^t_2+(\lambda_{\nu}-2\omega_2){\cal D}^t_3\right)u_{(3)}
=0,\\
\left((\lambda_{\nu}-2\omega_2){\cal D}_3+D_1{\cal D}_2\right)u_{(1)}
+\left(D^2_2-(\lambda_{\nu}-2\omega_2)
(\lambda_{\nu}-2\omega_3)\right)u_{(3)}=0.\\
\end{array}\right.
\end{equation}
\begin{Lemm}
\label{Ass}
If
\begin{equation}
\left(\lambda_{\nu}-2\omega_1\right)\left(\lambda_{\nu}-2\omega_2\right)
\neq D^2_1,
\end{equation}
then the following inequality holds
\begin{equation}
\left(\lambda_{\nu}-2\omega_2\right)\left(\lambda_{\nu}-2\omega_3\right)
\neq D^2_2.
\end{equation}
\end{Lemm}
{\bf Proof.} Observe, that if $\lambda_{\nu}=2\omega_2$, then
\begin{equation}
(\lambda_{\nu}-2\omega_1)(\lambda_{\nu}-2\omega_2)=0\neq D^2_1
\end{equation}
and 
\begin{equation}
(\lambda_{\nu}-2\omega_2)(\lambda_{\nu}-2\omega_1)=0\neq D^2_2.
\end{equation}
Therefore, it is left to consider the case when $\lambda_{\nu}\neq 2\omega_2$.
Assume that
\begin{equation}
\left(\lambda_{\nu}-2\omega_2\right)\left(\lambda_{\nu}-2\omega_3\right)= 
D^2_2.
\end{equation}
Then  from the second equation of~(\ref{Sys1})
\begin{equation}
\left((\lambda_{\nu}-2\omega_2){\cal D}_3+D_1{\cal D}_2\right)u_{(1)}=0.
\end{equation}
With respect to the standard scalar product 
$\langle \cdot,\cdot\rangle_{{\mathbb R}^m}$, $m=n,n+1$, we have:
\begin{equation}
0=\langle u_{(3)},\left((\lambda_{\nu}-2w_2){\cal D}_3+D_1{\cal D}_2\right)u_{(1)}
\rangle_{{\mathbb R}^n}=
\langle \left(D_1{\cal D}^t_2+(\lambda_{\nu}-2\omega_2){\cal D}^t_3\right)
u_{(3)},u_{(1)}\rangle_{{\mathbb R}^{n+1}}.
\end{equation}
Therefore, in order to satisfy the first equation of~(\ref{Sys1})
we necessarily have
\begin{equation}
\left\{
\begin{array}{l}
\left(D^2_1-(\lambda_{\nu}-2\omega_1)(\lambda_{\nu}-2\omega_2)\right)u_{(1)}=0,\\
\left(D_1{\cal D}^t_2+(\lambda_{\nu}-2\omega_2){\cal D}^t_3\right)u_{(3)}=0.\\
\end{array}\right.
\end{equation}
From here and under the assumption that
\begin{equation}
\left(\lambda_{\nu}-2\omega_1\right)\left(\lambda_{\nu}-2\omega_2\right)\neq D^2_1,
\end{equation}
we obtain that $u_{(1)}=0$. Because $D_j\neq 0$, $j=1,2,3$, the matrix
\begin{equation}
\left(D_1{\cal D}^t_2+(\lambda_{\nu}-2\omega_2){\cal D}^t_3\right)
\end{equation}
has rank $n$ and its kernel is $0$. Therefore, $u_{(3)}=0$. Finally, 
from~(\ref{u(2)}) we obtain that $u_{(2)}=0$ and thus, $\lambda_{\nu}$ is not
an eigenvalue for $H$. This verifies the lemma.


\

\noindent
Assume as in Lemma~\ref{Ass} that
\begin{equation}
\left(\lambda_{\nu}-2\omega_1\right)\left(\lambda_{\nu}-2\omega_2\right)\neq D^2_1,
\end{equation}
then from the first equation of~(\ref{Sys1}) we deduce that
\begin{equation}
\label{u(1)}
u_{(1)}=\frac{D_1{\cal D}^t_2+\left(\lambda_{\nu}-2\omega_2\right){\cal D}^t_3}{(\lambda_{\nu}-2\omega_1)(\lambda_{\nu}-2\omega_2)-D^2_1}u_{(3)}.
\end{equation}
Substituting~(\ref{u(1)}) into~(\ref{u(2)}) we obtain
\begin{equation}
\label{u(2')}
u_{(2)}=\frac{(\lambda_{\nu}-2\omega_1){\cal D}^t_2+D_1{\cal D}^t_3}{(\lambda_{\nu}-2\omega_1)(\lambda_{\nu}-2\omega_2)-D^2_1}u_{(3)}.
\end{equation}
Due to Lemma~\ref{Ass} 
\begin{equation}
\left(\lambda_{\nu}-2\omega_2\right)\left(\lambda_{\nu}-2\omega_3\right)\neq D^2_2.
\end{equation}
Thus, from the second equation of~(\ref{Sys1}) we have that
\begin{equation}
u_{(3)}=\frac{\left(\lambda_{\nu}-2\omega_2\right){\cal D}_3+D_1{\cal D}_2}{(\lambda_{\nu}-2\omega_2)(\lambda_{\nu}-2\omega_3)-D^2_2}u_{(1)}.
\end{equation}
Therefore, because $\lambda_{\nu}\neq 2\omega_2$, $u_{(3)}$ satisfies 
$$
\left((\lambda_{\nu}-2\omega_1)(\lambda_{\nu}-2\omega_2)(\lambda_{\nu}-2\omega_3)-(\lambda_{\nu}-2\omega_3)D^2_1-
(\lambda_{\nu}-2\omega_1)D^2_2-(\lambda_{\nu}-2\omega_2)D^2_3 \right)u_{(3)}=
$$
\begin{equation}
\label{Ker}
=
\left(D_1{\cal D}_3{\cal D}^t_2+D_1{\cal D}_2{\cal D}^t_3\right)u_{(3)}.
\end{equation}

\

\noindent
This leads us to the following theorem:


\begin{Th}
\label{Th3n+2}
For $j=1,...,n$ each of the three solutions of the cubic equation
$$
(\lambda_{\nu}-2\omega_1)(\lambda_{\nu}-2\omega_2)(\lambda_{\nu}-2\omega_3)-(\lambda_{\nu}-2\omega_2)D^2_3-
(\lambda_{\nu}-2\omega_1)D^2_2-(\lambda_{\nu}-2\omega_3)D^2_1=
$$
\begin{equation}
\label{MAIN}
=2D_1D_2D_3cos\left(\frac{\pi j}{n+1}\right)
\end{equation}
is an eigenvalue for $H$. The component $u_{(3)}$ of the corresponding eigenvector $u_{\nu}$ is
\begin{equation}
u_{(3)}=\left(sin\left(\frac{\pi j}{n+1}\right),sin\left(\frac{2\pi j}{n+1}\right),...,sin\left(\frac{n\pi j}{n+1}\right) \right).
\end{equation}
The remaining components $u_{(1)}$ and $u_{(2)}$ are determined uniquely from
\begin{equation}
u_{(1)}=\frac{\left(\lambda_{\nu}-2\omega_2\right){\cal D}^t_3+D_1{\cal D}^t_2}{(\lambda_{\nu}-2\omega_1)(\lambda_{\nu}-2\omega_2)-D^2_1}u_{(3)},
\end{equation}
and
\begin{equation}
u_{(2)}=\frac{(\lambda_{\nu}-2\omega_1){\cal D}^t_2+D_1{\cal D}^t_3}{(\lambda_{\nu}-2\omega_1)(\lambda_{\nu}-2\omega_2)-D^2_1}u_{(3)}.
\end{equation}
where ${\cal D}_2$, ${\cal D}_3$ are given in~(\ref{D2D3}).
\

\noindent
There are two other eigenvalues of $H$ which satisfy the following quadratic
equation
\begin{equation}
\left(\lambda_{\nu}-2\omega_1\right)\left(\lambda_{\nu}-2\omega_2\right)-D^2_1=0.
\end{equation}
The component $u_{(3)}$ of the corresponding eigenvector $u_{\nu}$ is 
necessarily zero, the component $u_{(1)}$ spans the (one-dimensional) kernel
of
\begin{equation}
(\lambda_{\nu}-2\omega_2){\cal D}_3+D_1{\cal D}_2
\end{equation}
and the component $u_{(2)}$ is
\begin{equation}
u_{(2)}=\frac{D_1}{\lambda_{\nu}-2\omega_2}\cdot u_{(1)}.
\end{equation}
All eigenvalues constructed are distinct and they exhaust all $3n+2$ distinct
eigenvalues for $H$.
\end{Th}
{\bf Proof.} 
Let us consider the case when $2w_2$ is an eigenvalue for $H$ and $u_{\nu}$ is 
the corresponding eigenvector. Then for the components $u_{(1)}$ 
and $u_{(3)}$ we have from the second equation of~(\ref{3n+2}):
\begin{equation}
u_{(1)}=-\frac{1}{D_1}\cdot {\cal D}^t_2u_{(3)}.
\end{equation}
From the first equation of~(\ref{3n+2}):
\begin{equation}
u_{(2)}=-\frac{1}{D_1}\left(\frac{2\omega_2-2\omega_1}{D_1}\cdot {\cal D}^t_2+
{\cal D}^t_3\right)u_{(3)}.
\end{equation}
Substituting this into the third equation of~(\ref{3n+2}) we obtain:
$$
\left(-\frac{1}{D_1}\cdot {\cal D}_3{\cal D}^t_2+
-\frac{1}{D_1}\left(\frac{2\omega_2-2\omega_1}{D_1}\cdot 
{\cal D}_2{\cal D}^t_2+
{\cal D}_2{\cal D}^t_3\right)
+(2\omega_3-2\omega_2)\right)u_{(3)}=
$$
\begin{equation}
=
\left((2\omega_3-2\omega_2)-(2\omega_2-2\omega_1)\frac{D^2_2}{D^2_1}-\frac{D_2D_3}{D_1}(J_n+J^t_n)
\right)
u_{(3)}=0.
\end{equation}
Thus, from Lemma~\ref{JLAMBDA}
\begin{equation}
2(\omega_3-\omega_2)D^2_1+2(\omega_1-\omega_2)D^2_2-2D_1D_2D_3cos\left(\frac{\pi j}{n+1}\right)=0
\end{equation}
for some $j=1,...,n$.
From here it follows that $2\omega_2$ also satisfies Eq.~(\ref{MAIN}), 
moreover, the formulae for the component $u_{(3)}$ and 
(as a consequence) for
$u_{(1)}$ and $u_{(2)}$ coincide with those given in the body of the theorem.
We obtain that the case $\lambda_{\nu}=2\omega_2$ can be considered 
simultaneously with all the other solutions of~(\ref{MAIN}).

Therefore, every eigenvalue for $H$ is either a solution for 
\begin{equation}
\left(\lambda_{\nu}-2\omega_1\right)\left(\lambda_{\nu}-2\omega_2\right)=D^2_1,
\end{equation}
or it is a 
solution of~(\ref{MAIN}) for some $j=1,...,n$. Because $H$ has exactly 
$3n+2$ eigenvalues and all of them are distinct (see Lemma~\ref{distinct}), 
all the solutions of the quadratic equation and of $n$ cubic equations are 
multiplicity free and pairwise distinct.

Now the first part of Theorem~\ref{Th3n+2} follows from the remarks above, 
from~(\ref{Ker}) and Lemma~\ref{JLAMBDA}. Indeed if~(\ref{Ker}) admits a 
nontrivial solution for some $\lambda_{\nu}$, then either $\lambda_{\nu}=
2\omega_2$
or  
\begin{equation}
\mu=\left((\lambda_{\nu}-2\omega_1)(\lambda_{\nu}-2\omega_2)
(\lambda_{\nu}-2\omega_3)-
(\lambda_{\nu}-2\omega_3)D^2_1-
(\lambda_{\nu}-2\omega_1)D^2_2-(\lambda_{\nu}-2\omega_2)D^2_3 \right)
\end{equation}
is an eigenvalue for $D_1{\cal D}_3{\cal D}^t_2+D_1{\cal D}_2{\cal D}^t_3$.
The first case has been already discussed, while for the second we 
apply Lemma~\ref{JLAMBDA}. It follows then that $\mu$ is necessarily of the 
form
\begin{equation}
2D_1D_2D_3cos\left(\frac{\pi j}{n+1}\right)
\end{equation}
for some $j=1,...,n$. As it was explained before, among those $3n$ eigenvalues 
constructed there are no repeated. The component $u_{(3)}$ of the 
corresponding eigenvectors are uniquely determined by the eigenvalues due to 
Lemma~\ref{JLAMBDA}. The other two components $u_{(1)}$ and $u_{(2)}$
are also uniquely determined by~(\ref{u(1)}) and~(\ref{u(2')}) respectively.
This gives us a unique eigenvector $u_{\nu}$ of $H$ for each solution 
of~(\ref{MAIN}).

The second part of the statement follows from~(\ref{Sys1}) because
\begin{equation}
\left(D_1{\cal D}^t_2+(\lambda_{\nu}-2\omega_2){\cal D}^t_3\right)
\end{equation}
has rank $n$ and its kernel is $0$.


\section{Multiple Quantum Spin Dynamics of an Inhomogeneous Spin Chain with
$3n+2$ Sites}

Information on the exact spectrum of the Hamiltonian of an open spin chain
provides us with the techniques for determining the multi-quantum 
dynamics in such a system. The MQ NMR dynamics of the nuclear spins coupled 
by the nearest neighbour dipolar interactions was developed in~\cite{FeldLac}. 
The corresponding Hamiltonian is
\begin{equation}
\label{MQ}
H_{MQ}=\frac{1}{2}\sum^{N-1}_{n=1}D_{n,n+1}\{I_{n,+}I_{n+1,+}+
I_{n,-} I_{n+1,-}\}.
\end{equation}
The Hamiltonian $H_{MQ}$~(\ref{MQ}) takes exactly the form of the 
Hamiltonian $H$~(\ref{TheorHam}) with the Larmor frequencies $\omega_n=0$ for 
all the sites, by making use of the unitary transformation~\cite{DorMakFel} 
\begin{equation}
Y=exp\left(-i\pi I_{2,x}\right)exp\left(-i\pi I_{4,x}\right)\cdots
exp\left(-i\pi I_{N-1,x}\right)
\end{equation}
acting on the even sites so that $YH_{MQ} Y^+=H$ (Eq.~(\ref{TheorHam});
$\{\omega_n=0\}$). The transformation $Y$ brings the initial density matrix
at the high temperature approximation (see~\cite{Abragam} for details):
\begin{equation}
\label{INI}
\rho(0)=\sum^N_{j=1}I_{j,z}=\sum^N_{j=1}\left(c^+_jc_j-1/2\right)
\end{equation}
to the form
\begin{equation}
\bar \rho(0)=YI_zY^+=\sum^N_{n=1}(-1)^{n-1}I_{n,z},
\end{equation}
where we introduce the total polarisation $I_z=\sum^N_{n=1} I_{n,z}$.
The Liouville--von Neumann equation ($\hbar=1$)
\begin{equation}
i\frac{\partial \rho}{\partial t}=[H_{MQ},\rho]
\end{equation}
with the Hamiltonian $H_{MQ}$~(\ref{MQ}) and the initial density 
matrix $\rho(0)$~(\ref{INI})
gives us the intensities $G_n(t)$ of $n=0$ and $n=\pm 2$ orders only with the 
conservation conditions~\cite{DorMakFel,LathHanGleas}
\begin{equation}
G_0(t)+G_2(t)+G_{-2}(t)=1.
\end{equation}

\begin{Th}
\label{Trace}
The intensity $G_n(t)$, $n=0,\pm 2$, of MQ coherences of the Hamiltonian 
$H_{MQ}$~(\ref{MQ}) are
\begin{equation}
\label{TrMQ}
G_0(t)=\frac{Tr \left[{\rm {cos}}^2\left(H_{MQ}\cdot t\right)\right]}{N},\quad 
G_{\pm 2}(t)=\frac{Tr \left[{\rm {sin}}^2\left(H_{MQ}\cdot t\right)\right]}{2N}.
\end{equation}
\end{Th}
{\bf Proof.} According to~\cite{FeldDor}
\begin{equation}
\label{GDF}
G_2(t)=G_{-2}(t)=\frac{1}{N}\sum_{k=1,3,...}\sum_{n=2,4,...}
\left|\sum^N_{j=1}(-1)^jS_{jk}S^*_{jn}\right|^2,
\end{equation}
with 
\begin{equation}
\label{EXPHMQ}
S_{jk}=\sum_l u^*_{jl}u_{kl}e^{-\frac{i}{2}\lambda_l t},
\end{equation}
where $\lambda_l$ are the eigenvalues of $H_{MQ}$, $l=1,...,N$, and 
the unitary matrix $U=\{u_{kl}\}^N_{k,l=1}$ diagonalises $H_{MQ}$. 
Let us rewrite~(\ref{GDF}) in the matrix form:
\begin{equation}
\label{MATR1}
G_2(t)=G_{-2}(t)=\frac{1}{N}
Tr\left(B_0AB_1A^*\right),
\end{equation}
where $B_0$ is the diagonal matrix with ones in odd rows and zeroes in 
even rows, $B_1$ is the diagonal matrix with ones in even rows and
zeroes in odd rows:
$$
B_0=diag\{1,0,1,0,...\},\quad B_1=diag\{0,1,0,1,...\};
$$
and the matrix $A$ is
\begin{equation}
A=S\left(B_1-B_0\right)S^*,\quad S=exp\left(-\frac{i}{2}H_{MQ}\cdot t\right).
\end{equation}
Observe that
\begin{equation}
\left(B_1-B_0\right)H_{MQ}=-H_{MQ}\left(B_1-B_0\right).
\end{equation}
Therefore,
\begin{equation}
\left(B_1-B_0\right)\cdot S\cdot \left(B_1-B_0\right)=S^*\quad
{\rm {and}}\quad
Tr\left[(S)^m\right]=Tr\left[\left(S^*\right)^m\right]. 
\end{equation}
Using
\begin{equation}
AA^*=A^2=I_N,\quad
A\left(B_1-B_0\right)A^*=\left(B_1-B_0\right)\left(S^*\right)^4
\end{equation}
and 
\begin{equation}
Tr\left[A\left(B_1-B_0\right)A^*\right]=
Tr\left[\left(B_1-B_0\right)A^*A\right]=
Tr\left[\left(B_1-B_0\right)AA^*\right]
\end{equation}
we deduce
$$
Tr\left[B_0AB_1A^*\right]=\frac{1}{4}Tr\left[\left(I-\left(B_1-B_0\right)\right)A
\left(I+\left(B_1-B_0\right)\right)A^*\right]=
$$ 
$$
=\frac{1}{4}Tr\left[AA^*-\left(B_1-B_0\right)AA^*+
A\left(B_1-B_0\right)A^*-\left(B_1-B_0\right)A\left(B_1-B_0\right)A^*\right]=
$$
$$ 
=\frac{1}{4}Tr\left[I_N-\left(S^*\right)^4\right]=
-\frac{1}{8}Tr\left[S^4+\left(S^*\right)^4-2I_N\right]=
$$
\begin{equation}
=\frac{1}{2}Tr\left[\left(\frac{S^2-\left(S^*\right)^2}{2i}\right)^2\right]
=\frac{1}{2}Tr\left[{\rm {sin}}^2\left(H_{MQ}\cdot t\right)\right].
\end{equation}
From here the result follows.

\begin{Rem}
In~\cite{FeldmanRud} Formula~(\ref{TrMQ}) was proposed for alternating
spin chains. Theorem~\ref{Trace} establishes the same formula for arbitrary
spin chains coupled by the nearest neighbour dipolar interactions.  
\end{Rem}

\

Theorem~\ref{Trace} together with Theorem~\ref{Th3n+2} allow us to 
calculate 
MQ coherence intensities of the zero and the second orders, $G_0(t)$ and 
$G_2(t)+G_{-2}(t)$, without 
performing matrix multiplication, by solving $O(N)$ cubic 
equations (compare with~\cite{FeldDor}). Let us consider a linear spin chain of
length $1001$ which consists of four-spin fragments represented in Figure 1.

\begin{picture}(150,80)
\put(30,50){\circle*{16}}
\put(27,30){$1$}
\put(38,50){\line(1,0){50}}
\put(50,54){$D_{12}$}
\put(96,50){\circle*{16}}
\put(94,30){$2$}
\put(104,50){\line(1,0){70}}
\put(125,54){$D_{23}$}
\put(182,50){\circle*{16}}
\put(182,30){$3$}
\put(190,50){\line(1,0){90}}
\put(230,54){$D_{34}$}
\put(288,50){\circle*{16}}
\put(288,30){$4$}
\put(296,50){\line(1,0){50}}
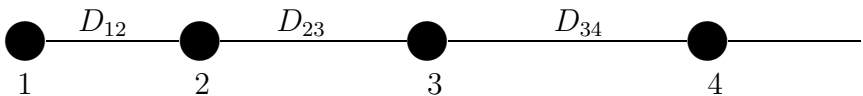
\put(17,4){\footnotesize{Figure 1. The four--spin fragment of the linear 
chain. The distances between}} 
\put(9,-7){\footnotesize{neighbouring spins in the fragments are 2.7, 3 
and 3.3 $\dot A$. The dipolar coupling }}
\put(10,-18){\footnotesize{constants are $D_{12}=2\pi 6096$,
$D_{23}=2\pi 4444$, $D_{34}=2\pi 3339s^{-1}$.}}
\end{picture} 

\vspace{2cm}

The distances between neighbouring spins in the fragments are $2.7\dot A$,
$3 \dot A$ and $3.3 \dot A$. The dipolar coupling constants 
$D_{12}=2\pi 6096$, $D_{23}=2\pi 4444$, $D_{34}=2\pi 3339 s^{-1}$ are
used in all numerical calculations. The intensities of MQ coherences for the
inhomogeneous linear chain with $N=1001$ spins consisting of fragments of
Figure 1 are shown in Figure 2. The dynamics behaviour coincides with the
one given in~\cite{FeldDor}.  



\section{The Generalisation of the Method for Spin Chains with {\bf $kn-1$} 
Sites}

Now we consider the case of a $k$-periodic system with $kn-1$-sites. 
According to Section 3, to diagonalise Hamiltonian of such a system
we have to solve the following system of linear equations
\begin{equation}
{\mathrm {\bf H}}
\left[
\begin{array}{c}
u_{(1)}\\
\vdots\\
u_{(k)}\\
\end{array}\right]=
\lambda_{\nu}
\left[
\begin{array}{c}
u_{(1)}\\
\vdots\\
u_{(k)}\\
\end{array}\right],
\end{equation}
where
\begin{equation}
{\mathrm {\bf H}}=
\left[\begin{array}{cccccc}
& & & & & {\cal D}^t_k\\
& & & & &    0\\
& & {\cal H}_{1,k-1}& &  & \vdots\\
& &  & &  & 0\\
& &  &      &  &    {\cal D}^t_{k-1}\\
{\cal D}_k&0 &\cdots &0 & {\cal D}_{k-1} &2{\cal W}_k\\
\end{array}\right],
\end{equation}
and 
\begin{equation}
{\cal H}_{1,k-1}=
\left[\begin{array}{cccccc}
2{\cal W}_1   & {\cal D}_1 & 0   &\cdots&0&0\\
{\cal D}_1 & 2{\cal W}_{2}   & {\cal D}_{2} &\cdots&0&0\\
0   & {\cal D}_{2} & 2{\cal W}_{3}   &\cdots&0&0\\
\vdots&\vdots&\vdots&\ddots&\vdots&\vdots\\
0&0&0&\cdots&2{\cal W}_{k-2} &{\cal D}_{k-2}\\
0&0&0&\cdots&{\cal D}_{k-2}&2{\cal W}_{k-1}\\
\end{array}\right].
\end{equation}
Observe that because all ${\cal D}_j$, $j=1,...,k-2$, are diagonal
$n\times n$ matrices, it is also true that
\begin{equation}
{\cal H}_{1,k-1}=H_{1,k-1}\otimes I_n.
\end{equation}
Assume $\lambda_{\nu}$ is not an eigenvalue for ${\cal H}_{1,k-1}$
(equivalently, is not an eigenvalue for $H_{1,k-1}$),
and consider the matrix
\begin{equation}
{\mathrm {\bf G}}=\left[\begin{array}{cccc}
                                         &                  & &0\\
&\left({\cal H}_{1,k-1}-\lambda_{\nu}I_{n(k-2)}\right)^{-1} & &\vdots\\
                                         &                  & &0\\
0&\cdots  &0  &I_{n-1}\\
\end{array}\right].
\end{equation}
Under the above assumption on $\lambda_{\nu}$ the eigenvector equation
\begin{equation}
\label{TRINEW}
{\mathrm {\bf H}}\left[\begin{array}{c}
u_{(1)}\\
\vdots\\
u_{(k)}\\
\end{array}\right]=\lambda_{\nu}\left[\begin{array}{c}
u_{(1)}\\
\vdots\\
u_{(k)}\\
\end{array}\right]
\end{equation}
is equivalent to
\begin{equation}
{\mathrm {\bf G}}\left({\mathrm {\bf H}}-\lambda_{\nu}
I_{nk-1}\right)\left[\begin{array}{c}
u_{(1)}\\
\vdots\\
u_{(k)}\\
\end{array}\right]=0,
\end{equation}
or in another form to
\begin{equation}
\left[
\begin{array}{cccccc}
1 &  0&\cdots & 0& 0\\
0 &  1&\cdots & 0& 0\\
\vdots & \ddots&\ddots&\ddots& \vdots\\
0 &  0& \cdots & 1& 0\\
0 &  0& \cdots & 0& 1\\
{\cal D}_k &  0& \cdots & 0& {\cal D}_{k-1}\\
\end{array}
{\mathrm {\bf G}}\left[\begin{array}{c}
{\cal D}^t_k\\
0\\
\vdots\\
0\\
{\cal D}^t_{k-1}\\
2{\cal W}_k-\lambda_{\nu}I_{n-1}\\
\end{array}\right]\right]
\left[\begin{array}{c}
u_{(1)}\\
u_{(2)}\\
\vdots\\
u_{(k-2)}\\
u_{(k-1)}\\
u_{(k)}\\
\end{array}\right]=0.
\end{equation}
Let us denote elements of the matrix 
$\left(H_{1,k-1}-\lambda_{\nu} I_{k-1}\right)^{-1}$ by $P_{i,j}$ 
($i$ --- row, $j$ --- column).
Using
$$
\left({\cal H}_{1,k-1}-\lambda_{\nu}I_{n(k-1)}\right)^{-1}=
\left(H_{1,k-1}\otimes I_n-\lambda_{\nu}I_{k-1}\otimes I_n\right)^{-1}=
\left(H_{1,k-1}-\lambda_{\nu}I_{k-1}\right)^{-1}\otimes I_n,
$$
we derive from~(\ref{TRINEW}) for $k-1\ge j\ge 1$ that
\begin{equation}
\label{NEWu_j}
u_{(j)}=
-\left(P_{j,1}\cdot {\cal D}^t_k+P_{j,k-1}\cdot {\cal D}^t_{k-1}\right)u_{(k)}.
\end{equation}
Substituting this expression with $j=1$ and $j=k-1$ into the last equation 
of~(\ref{TRINEW}) we obtain:
$$
\left(P_{1,1}\cdot {\cal D}_k
{\cal D}^t_k+P_{1,k-1}{\cal D}_k{\cal D}^t_{k-1}+
P_{k-1,1}\cdot {\cal D}_{k-1}{\cal D}^t_k+P_{k-1,k-1}\cdot 
{\cal D}_{k-1}{\cal D}^t_{k-1}\right)u_{(k)}=
$$
\begin{equation}
=
\left(2\omega_k-\lambda_{\nu}\right)u_{(k)}.
\end{equation}
Because
$$
{\cal D}_{k-1}{\cal D}^t_{k-1}=D^2_{k-1} I_{n-1},\quad
{\cal D}_k{\cal D}^t_k=D^2_k I_{n-1},\quad
{\cal D}_{k-1}{\cal D}^t_k=D_{k-1}D_k J^t_{n-1},
$$
\begin{equation}
{\cal D}_k{\cal D}^t_{k-1}=D_{k-1}D_k J_{n-1},
\end{equation}
with the help of Corollary~\ref{INVERSE} we derive that if $\lambda_{\nu}$ is
not an eigenvalue of $H_{1,k-1}$, then the component $u_{(k)}$ of the 
corresponding eigenvector $u_{\nu}$ belongs to the kernel of
$$
{\cal M}=\left(det\left(H_{2,k-1}-\lambda_{\nu} I_{k-2}\right)D^2_k+
det\left(H_{1,k-2}-\lambda_{\nu} I_{k-2}\right)D^2_{k-1}-
\right.
$$
\begin{equation}
\left.
-(2\omega_k-\lambda_{\nu}) det\left(H_{1,k-1}-\lambda_{\nu} I_{k-2}\right)
\right)\cdot I_{n-1}
+
(-1)^kD_1\cdots D_k \left(J_{n-1}+J^t_{n-1}\right).
\end{equation}
Since
$$
(2\omega_k-\lambda_{\nu}) det\left(H_{1,k-1}-\lambda_{\nu} I_{k-2}\right)-
det\left(H_{1,k-2}-\lambda_{\nu} I_{k-2}\right)D^2_{k-1}=
det\left(H_{1,k}-\lambda_{\nu}I_k\right)
$$
we can simplify the expression for ${\cal M}$:
$$
{\cal M}=
\left(det\left(H_{2,k-1}-\lambda_{\nu} I_{k-2}\right)D^2_k-
det\left(H_{1,k}-\lambda_{\nu} I_{k}\right)D^2_{k-1}\right)\cdot I_{n-1}
+
$$
\begin{equation}
+
(-1)^kD_1\cdots D_k \left(J_{n-1}+J^t_{n-1}\right).
\end{equation}
Therefore,
$$
\left(
det\left(H_{1,k}-\lambda_{\nu} I_{k}\right)D^2_{k-1}
-det\left(H_{2,k-1}-\lambda_{\nu} I_{k-2}\right)D^2_k
\right)u_{(k)}=
$$
\begin{equation}
\label{MAINEQ}
=(-1)^kD_1\cdots D_k \left(J_{n-1}+J^t_{n-1}\right)u_{(k)}.
\end{equation}
We are ready to state the main result of the paper.
\begin{Th}
\label{MAINkn-1}
Each eigenvalue of the Hamiltonian $H$ of a $k$-periodic system with 
$kn-1$ sites is either an eigenvalue of $H_{1,k-1}$ or it is a solution
of the equation
\begin{equation}
\label{EQU}
det\left(H_{1,k}-\lambda_{\nu} I_k\right)
-det\left(H_{2,k-1}-\lambda_{\nu} I_{k-2}\right)D^2_k=
(-1)^k 2D_1\cdots D_k cos\left(\frac{\pi j}{n}\right),
\end{equation}
for some $j=1,...,n-1$. Equation~(\ref{EQU}) does not have repeated roots and 
all $k(n-1)$ solutions constructed  from~(\ref{EQU}) are pairwise distinct and 
are not eigenvalues for $H_{1,k-1}$.

If $\lambda_{\nu}$ is the solution of~(\ref{EQU}) for some $j=1,...,n-1$, then 
it is an eigenvalue for $H$ and the component $u_{(k)}$ of the corresponding 
eigenvector $u_{\nu}$ is
\begin{equation}
u_{(k)}=\left(sin\left(\frac{\pi j}{n}\right),\dots,
sin\left(\frac{(n-1)\pi j}{n}\right)\right).
\end{equation}
The other components $u_{(j)}$, $j=1,...,k-1$, are determined uniquely from
$$
u_{(j)}=\frac{(-1)^{j-1}}{det\left(H_{1,k-1}-\lambda_{\nu}I_{k-1}\right)}\cdot
\left[D_1\cdots D_{j-1}
det\left(H_{j+1,k-1}-\lambda_{\nu}I_{k-j-1}\right)\cdot {\cal D}^t_k+
\right.
$$
\begin{equation}
\left.
+(-1)^{k-1}det\left(H_{1,j-1}-\lambda_{\nu}I_{j-1}\right)
D_j\cdots D_{k-2}{\cal D}^t_{k-1}\right]u_{(k)},
\end{equation}
where ${\cal D}_{k-1}$, ${\cal D}_k$ are given in~(\ref{DdDk}) ($d=k-1$).
Every eigenvalue $\lambda_{\nu}$ of $H_{1,k-1}$ is an eigenvalue of $H$.   
The component $u_{(k)}$ of the corresponding 
eigenvector $u_{\nu}$ of $H$ is zero. The component $u_{(1)}$ spans the 
one--dimensional kernel of
\begin{equation}
(-1)^{k-1}D_1\cdots D_{k-2}{\cal D}_{k-1}-
det\left(H_{2,k-1}-\lambda_{\nu}I_{k-2}\right){\cal D}_k.
\end{equation}
The remaining components $u_{(j)}$, $j=2,...,k-1$, are
\begin{equation}
\label{THEOTHER}
u_{(j)}=(-1)^{j-1}\frac{D_1\cdots D_{j-1}
det\left(H_{j+1,k-1}-\lambda_{\nu}I_{k-j-1}\right)}{det\left(H_{2,k-1}-\lambda_{\nu}I_{k-2}\right)}u_{(1)}.
\end{equation}
\end{Th}
{\bf Proof.}
The first part of the theorem follows from Lemma~\ref{JLAMBDA} because $H$ has 
exactly $kn-1$ distinct eigenvalues. 
Indeed, if an eigenvalue $\lambda_{\nu}$ of $H$ is not an eigenvalue of 
$H_{1,k-1}$, then as it was shown above, the component $u_{(k)}$ of the 
corresponding eigenvector $u_{\nu}$ of $H$ satisfies~(\ref{MAINEQ}).
Therefore, $\lambda_{\nu}$ satisfies~(\ref{EQU}) for some $j=1,...,n-1$.

For the second part of the theorem we again use Lemma~\ref{JLAMBDA}. The 
component $u_{(k)}$ of the corresponding eigenvector $u_{\nu}$, 
satisfies~(\ref{MAINEQ}), and, therefore, is uniquely determined by 
$\lambda_{\nu}$ as stated in Lemma~\ref{JLAMBDA}. The remaining components of 
the eigenvectors $u_{(\nu)}$ corresponding to the solutions of~(\ref{EQU}) can 
be reconstructed from~(\ref{NEWu_j}) using Corollary~\ref{INVERSE}.

Finally, for the last part we observe that from Lemma~\ref{adj} 
and the property of the adjoint matrix
$$
D_{j-1}(-1)^{j-2}\frac{D_1\cdots D_{j-2}
det\left(H_{j,k-1}-\lambda_{\nu}I_{k-j}\right)}{det\left(H_{2,k-1}-\lambda_{\nu}I_{k-2}\right)}
+
$$
$$
+
\left(2\omega_j-\lambda_{\nu}\right) (-1)^{j-1}\frac{D_1\cdots D_{j-1}
det\left(H_{j+1,k-1}-\lambda_{\nu}I_{k-j-1}\right)}{det\left(H_{2,k-1}-\lambda_{\nu}I_{k-2}\right)}
+
$$
$$
+D_j (-1)^{j}\frac{D_1\cdots D_{j}
det\left(H_{j+2,k-1}-\lambda_{\nu}I_{k-j-2}\right)}
{det\left(H_{2,k-1}-\lambda_{\nu}I_{k-2}\right)}=
$$
$$
=D_{j-1}\cdot \frac{\mathrm {adj}_{1,j-1}
\left\{\left(H_{1,k-1}-\lambda_{\nu}I_{k-1}\right)\right\}}
{det\left(H_{2,k-1}-\lambda_{\nu} I_{k-2}\right)}+
\left(2\omega_j-\lambda_{\nu}\right)\cdot \frac{\mathrm {adj}_{1,j}
\left\{\left(H_{1,k-1}-\lambda_{\nu}I_{k-1}\right)\right\}}
{det\left(H_{2,k-1}-\lambda_{\nu} I_{k-2}\right)}+
$$
\begin{equation}
+D_j\cdot \frac{\mathrm {adj}_{1,j+1}
\left\{\left(H_{1,k-1}-\lambda_{\nu}I_{k-1}\right)\right\}}
{det\left(H_{2,k-1}-\lambda_{\nu} I_{k-2}\right)}=\delta_{1,j}
\frac{det\left(H_{1,k-1}-\lambda_{\nu} I_{k-1}\right)}
{det\left(H_{2,k-1}-\lambda_{\nu} I_{k-2}\right)}=0,
\end{equation}
where $k-1>j>1$ and $\delta_{i,j}$ is the Kronecker symbol. We also have
\begin{equation}
D_{k-2}(-1)^{k-3}\frac{D_1\cdots D_{k-3}
\left(2\omega_{k-1}-\lambda_{\nu}\right)}{det\left(H_{2,k-1}-\lambda_{\nu}I_{k-2}\right)}
+\left(2\omega_{k-1}-\lambda_{\nu}\right)(-1)^{k-2}\frac{D_1\cdots D_{k-2}}
{det\left(H_{2,k-1}-\lambda_{\nu}I_{k-2}\right)}=0.
\end{equation}
If $\lambda_{\nu}$ is an eigenvalue 
for $H_{1,k-1}$, then
\begin{equation}
2\omega_1+D_1(-1)\frac{D_1det\left(H_{3,k-1}-\lambda_{\nu}I_{k-2}\right)}{det\left(H_{2,k-1}-\lambda_{\nu}I_{k-2}\right)}
=\frac{det\left(H_{1,k-1}-\lambda_{\nu}I_{k-2}\right)}
{det\left(H_{2,k-1}-\lambda_{\nu}I_{k-2}\right)}=0.
\end{equation}
Therefore, if $u_{(1)}$ spans the kernel of
\begin{equation}
(-1)^{k-1}D_1\cdots D_{k-2}{\cal D}_{k-1}-
det\left(H_{2,k-1}-\lambda_{\nu}I_{k-2}\right){\cal D}_k
\end{equation}
and $u_{(k)}=0$, then vector $u_{\nu}$ with the other components
given by~(\ref{THEOTHER}) does satisfy the eigenvalue equation
\begin{equation}
\left(H-\lambda I_{kn-1}\right)u_{\nu}=0.
\end{equation}
This completes the proof.

\section{Conclusion}

In the paper we proposed a new one-dimensional exactly solvable  model for  
a linear $k$-periodic (in space) open spin chain with $kn-1$ sites. For the 
diagonalisation 
procedure it is important that the number of sites is $(k-1)(\bmod k)$. 
Nevertheless, the model can serve as a good approximation for any linear 
periodic spin system if the number of sites in it is much more than the period.

The developed method of diagonalisation of the XY-Hamiltonian of
inhomogeneous linear spin chains can be applied to different problems of 
quantum information theory [3,7,8] and spin dynamics. This method allows us 
to avoid matrix multiplications which are time consuming operations in the
systems with large numbers of spins. In some cases we can suggest analytical
methods for problems of spin dynamics instead of the known numerical ones [9].

The  suggested method could also be applied to different physical and
technical problems which use three--diagonal matrices~\cite{Il'in}. In 
particular, new numerical methods of solving such problems could be worked 
out on the basis of the approach proposed in the paper.

\section{Acknowledgments}
The author has been supported by EPSRC grant GR/S92137/01.


\end{document}